\documentclass[preprint2,usenatbib,twocolappendix]{aastex631}
\usepackage{lipsum}
\usepackage{totcount}
\usepackage{comment}
\usepackage{amsmath}
\newtotcounter{citnum} 
\def\oldbibitem{} \let\oldbibitem=\bibitem
\def\bibitem{\stepcounter{citnum}\oldbibitem}
\usepackage{array}
\usepackage{floatrow}
\usepackage[caption=false]{subfig}
\newcommand{\citewm}{\citet{PowellSulphurDioxideMidinfrared2024}}

\graphicspath{{./}{figures/}}

\shorttitle{Debris Disk around WASP-39?}
\shortauthors{Flagg et al.}

\begin{document}

\title{Debris Disks can Contaminate Mid-Infrared Exoplanet Spectra: Evidence for a Circumstellar Debris Disk around Exoplanet Host WASP-39}


\correspondingauthor{Laura Flagg}
\email{laura.s.flagg@gmail.com}

\author[0000-0001-6362-0571]{Laura Flagg}
\affiliation{Department of Astronomy and Carl Sagan Institute, Cornell University, Ithaca, New York 14853, USA}

\author[0000-0001-6654-7859]{Alycia J. Weinberger}
\affiliation{Earth and Planets Laboratory, Carnegie Institution for Science, Washington, DC 20015, USA}

 \author[0000-0003-4177-2149]{Taylor J. Bell}
 \affiliation{Bay Area Environmental Research Institute, NASA's Ames Research Center, Moffett Field, CA 94035, USA}
 \affiliation{Space Science and Astrobiology Division, NASA's Ames Research Center, Moffett Field, CA 94035, USA}

 \author[0000-0003-0156-4564]{Luis Welbanks}
 \thanks{NHFP Sagan Fellow}
 \affiliation{School of Earth \& Space Exploration, Arizona State University, Tempe, AZ, 85257, USA}

\author[0000-0002-4262-5661]{Giuseppe Morello}
\affiliation{Department of Space, Earth and Environment, Chalmers University of Technology, SE-412 96 Gothenburg, Sweden}
\affiliation{Instituto de Astrof\'isica de Canarias (IAC), 38205 La Laguna, Tenerife, Spain}
\affiliation{Instituto de Astrof\'isica de Andaluc\'ia (IAA-CSIC), Glorieta de la Astronom\'ia s/n, 18008 Granada, Spain}

\author[0000-0002-4250-0957]{Diana Powell}
\affiliation{Center for Astrophysics ${\rm \mid}$ Harvard {\rm \&} Smithsonian, Cambridge, USA}

\author[0000-0003-4733-6532]{Jacob L. Bean}
\affiliation{Department of Astronomy \& Astrophysics, University of Chicago}

\author[0000-0002-0769-9614]{Jasmina Blecic}
\affiliation{Department of Physics, New York University Abu Dhabi, Abu Dhabi, UAE}
\affiliation{Center for Astro, Particle and Planetary Physics (CAP3), New York University Abu Dhabi, Abu Dhabi, UAE}

\author[0000-0001-7866-8738]{Nicolas Crouzet}
\affiliation{Leiden Observatory, Leiden University, P.O. Box 9513, 2300 RA Leiden, The Netherlands}

\author[0000-0002-8518-9601]{Peter Gao}
\affiliation{Earth and Planets Laboratory, Carnegie Institution for Science, Washington, DC 20015, USA}

\author[0000-0001-9164-7966]{Julie Inglis}
\affiliation{Division of Geological and Planetary Sciences, California Institute of Technology, Pasadena, CA, 91125}

\author[0000-0002-4207-6615]{James Kirk}
\affiliation{Imperial College London}

\author[0000-0003-3204-8183]{Mercedes López-Morales}
\affiliation{Center for Astrophysics ${\rm \mid}$ Harvard {\rm \&} Smithsonian, Cambridge, USA}

\author[0000-0002-0502-0428]{Karan Molaverdikhani}
\affiliation{Universitäts-Sternwarte, Ludwig-Maximilians-Universität München, Scheinerstrasse 1, D-81679 München, Germany}
\affiliation{Exzellenzcluster Origins, Boltzmannstraße 2, 85748 Garching, Germany}

\author[0000-0002-6500-3574]{Nikolay Nikolov}
\affiliation{Space Telescope Science Institute}

\author[0000-0002-1655-0715]{Apurva V. Oza}
\affiliation{Jet Propulsion Laboratory, California Institute of Technology, Pasadena, USA}

\author[0000-0002-3627-1676]{Benjamin V. Rackham}
\affiliation{Department of Earth, Atmospheric and Planetary Sciences, Massachusetts Institute of Technology, 77 Massachusetts Avenue, Cambridge, MA 02139, USA}
\affiliation{Kavli Institute for Astrophysics and Space Research, Massachusetts Institute of Technology, Cambridge, MA 02139, USA}

\author[0000-0003-3786-3486]{Seth Redfield}
\affiliation{Astronomy Department and Van Vleck Observatory, Wesleyan University, Middletown, CT 06459, USA}

\author[0000-0002-8163-4608]{Shang-Min Tsai}
\affiliation{University of California, Riverside}

\author[0000-0001-5349-6853]{Ray Jayawardhana}
\affiliation{Department of Astronomy, Cornell University, Ithaca, New York 14853, USA}
\affiliation{Department of Earth \& Planetary Sciences, Johns Hopkins University,  Baltimore, MD, 21218, USA}

\author[0000-0003-0514-1147]{Laura Kreidberg}
\affiliation{Max Planck Institute for Astronomy}

\author[0000-0001-8236-5553]{Matthew C. Nixon}
\affiliation{Department of Astronomy, University of Maryland, College Park}

\author[0000-0002-7352-7941]{Kevin B. Stevenson}
\affiliation{Johns Hopkins APL, Laurel, MD, 20723, USA}

\author[0000-0001-7836-1787]{Jake D. Turner}
\affiliation{Department of Astronomy and Carl Sagan Institute, Cornell University, Ithaca, New York 14853, USA}
 \thanks{NHFP Sagan Fellow}





\begin{abstract}

The signal from a transiting planet can be diluted by astrophysical contamination.  In the case of circumstellar debris disks, this contamination could start in the mid-infrared and vary as a function of wavelength, which would then change the observed transmission spectrum for any planet in the system.  The MIRI/LRS WASP-39b transmission spectrum shows an unexplained dip starting at $\sim$10 $\mu$m that could be caused by astrophysical contamination.   The spectral energy distribution displays excess flux at similar levels to that which are needed to create the dip in the transmission spectrum.  In this article, we show that this dip is consistent with the presence of a bright circumstellar debris disk, at a distance of $>$2 au. We discuss how a circumstellar debris disk like that could affect the atmosphere of WASP-39b.  We also show that even faint debris disks can be a source of contamination in MIRI exoplanet spectra.

\end{abstract}

\section{Introduction}
Transit spectroscopy is one of the best ways to study atmospheres of exoplanets. By measuring how much light from the star the planet blocks as a function of wavelength, we can learn important information about planets, such as their atmospheric composition \citep[e.g.][]{KreidbergExoplanetAtmosphereMeasurements2018}.

However, transit spectroscopy has the same weaknesses as single-band transit data. One major weakness is that blended targets can result in a shallower transit depth. The most common blends are stellar companions (e.g. \citealp{DamianoNearIRTransmissionSpectrum2017,EdwardsARESWASP76Tale2020}), and emission from the planet itself \citep{KippingNightsidePollutionExoplanet2010,MorelloPhasecurvePollutionExoplanet2021}. If the contaminant spectrum differs from that of the target star, it may imprint features on the transit spectrum and bias our interpretation of the data \citep[e.g.][]{EdwardsARESWASP76Tale2020}. 

Another common source of flux from planetary systems are circumstellar debris disks.  Debris disks are the remains from planet formation, akin to our own Asteroid Belt or Kuiper Belt.  Collisions of the planetesimals in these belts produce dust that emits in the infrared (IR) and millimeter wavelengths. This is evident from the spectral energy distribution (SED) covering those wavelengths \citep[and sources therein]{WyattEvolutionDebrisDisks2008}.   While dust levels from the Asteroid Belt or Kuiper Belt are not yet detectable in other Solar Systems \citep{HughesDebrisDisksStructure2018}, larger amounts of dust are frequently seen in other stellar and planetary systems.  While often quite faint, disks can be the dominant source of flux at mid and far IR wavelengths \citep{ChenSpitzerInfraredSpectrograph2014}.  In particular, small silicate particles cause a well-studied emission feature at 10 $\mu$m \citep[and sources therein]{HenningCosmicSilicates2010}.  At high enough flux ratios, which are plausible for debris disks, this extra emission could contaminate transit spectra of orbiting exoplanets.

WASP-39b is a hot Saturn around a main-sequence G8 star and a target of the JWST Transiting Exoplanet Early Release Science Program (JWST-DD-1366) and an ancillary DDT program (JWST-DD-2783). As such, its atmosphere has been extensively characterized through transmission spectroscopy with 4 different instruments over a wavelength range between 0.6 and 12 $\mu$m \citep[]{RustamkulovEarlyReleaseScience2023, AldersonEarlyReleaseScience2023, FeinsteinEarlyReleaseScience2023, AhrerEarlyReleaseScience2023, PowellSulphurDioxideMidinfrared2024,Welbanks2024}.  Analysis from the spectra show the planet has a low C/O ratio and a high metallicity.  The transmission spectrum also shows a mysterious (and  unexplained) dip longward of 10 $\mu$m \citep{PowellSulphurDioxideMidinfrared2024}.

In this paper, we look at the case of the MIRI transit spectrum of WASP-39b and show that this dip could plausibly be caused by a circumstellar debris disk around the host star diluting the transit at wavelengths longer than 10 $\mu$m. We show that the system's SED is also consistent with a star surrounded by a debris disk. We model the debris disk and discuss the range of parameters a debris disk could have.  We also discuss how a debris disk could affect the atmosphere of WASP-39b.

\section{WASP-39 MIRI-LRS Data}
The 5-12 $\mu$m transmission spectrum of WASP-39b (Director’s Discretionary Time, PID: 2783) was observed using JWST MIRI/LRS \citep{KendrewMidInfraredInstrumentJames2015} on 2023-02-14  to confirm the presence of atmospheric SO$_2$
 The details of the observations and analysis were presented in \citewm.

\section{Methods}

\subsection{Modelling the Stellar SED}\label{sec:stellarsed}

To model the stellar flux in the IR, we used photometry from WISE \citep{WrightWidefieldInfraredSurvey2010}, in addition to the out-of-transit stellar spectra taken for the transit observations.  For the purposes of evaluating the stellar flux, we only used the post-transit observations in order to minimize the effect of the ramp seen at the beginning of MIRI time series observations.  We used the \textit{rateints} files created from the Eureka pipeline for \citewm; those files were produced using version 0.9 of the \texttt{Eureka!} \citep{BellEurekaEndtoEndPipeline2022} pipeline, CRDS version 11.16.16 and context 1045, \texttt{jwst} package version 1.8.3 \citep{BushouseJWSTCalibrationPipelineVersion182_2022}. During the Stage 1 processing, the jump rejection threshold was increased to 7 and the \texttt{lastframe} step was also applied to remove the excessively noisy last frame from each integration. 

\begin{figure}[htb!]
\includegraphics[width=0.98\textwidth]{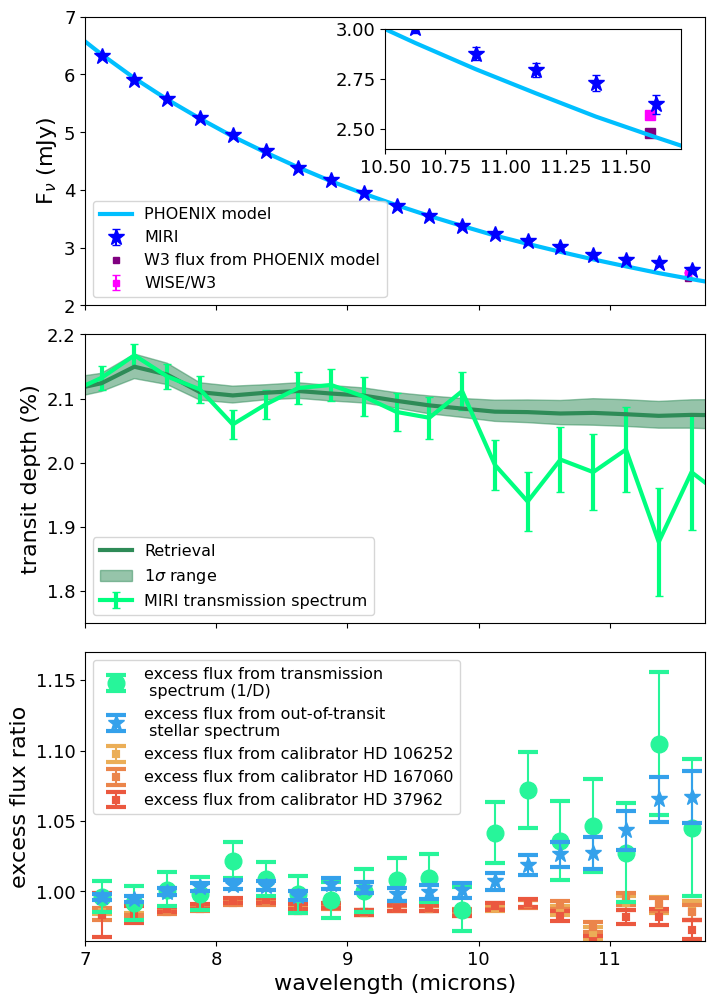}
\caption{(top) A portion of the observed flux from the  WASP-39 system along with a BT-Settl photospheric model, using WISE-W2 as an anchor point.  Both the MIRI data, as well as the W3 point from WISE hint at a slight amount of excess flux.  (middle) The WASP-39b MIRI transmission spectrum along with the best-fit model retrieved to the data short of 10 $\mu$m, along with 1$\sigma$ uncertainties. (bottom) The excess flux calculated using the stellar flux  compared with a photospheric model flux from the top plot (in blue) and the excess flux calculated from the transmission spectrum (in green). We also plot the excess stellar flux from three G star calibrators. Note: the uncertainties are the 1$\sigma$, random uncertainties calculated from the scatter in the data.  Systematic uncertainties are not included.  In the case of the flux-calibrated stellar spectrum, the systematic uncertainties may be on the order of 10\%.}
\label{fig:stellar}
\end{figure}

From there, we ran the JWST pipeline, version 1.11.0 \citep{BushouseJWSTCalibrationPipelineVersion111_2023}, with the following modifications: we included the flux calibration and the \textit{pixel\_replace} step to deal with bad pixels, and for the extraction, we used a tapered profile that is the width of 3 times the full width at half maximum.  Our reference file is at \doi{10.5281/zenodo.8423535}. We derive uncertainties by calculating the scatter between integrations using the JWST pipeline.  The data was then binned to match the resolution of the transmission spectrum from \citewm.  The resulting spectrum is plotted in Figure \ref{fig:stellar}, top.  We then compared the observed flux from MIRI and WISE/W3 to what we would expect from the photosphere of a star alone, using a PHOENIX/BT-Settl photospheric model at 5400 K, log(g)=4.5 at solar metallicity \citep{AllardModelAtmospheresSpectra2003,AllardModelsVerylowmassStars2012}\footnote{https://archive.stsci.edu/hlsps/reference-atlases/cdbs/grid/phoenix/}.  We  scaled the model to a distance of 213.3 pc \citep{Bailer-JonesEstimatingDistancesParallaxes2021}, assuming a radius of 0.9 R$_\odot$, and then by a factor of 0.956 to match the flux measured  by WISE's W2 filter (centered at $\sim$4.6 $\mu$m) using \texttt{synphot}. The photospheric model matches the MIRI data well between 7 and 10 $mu$m validating our choice to scale the model to the W2 flux and showing that the flux calibration of the MIRI data at those wavelengths is reasonable. 
Given we are in the Rayleigh-Jeans tail of the spectrum, we assume the photospheric model is accurate and do not propagate any potential uncertainties in the model flux from assumed stellar properties or the W2 flux when comparing the model flux to the observed MIRI data. The W3 bandpass goes from $\sim$8 to 15 $\mu$m, while the MIRI data ends at 12 $\mu$m, so we cannot directly compare the MIRI data to the W3 photometry. In the top panel of Figure \ref{fig:stellar}, we see hints of excess flux past 10 $\mu$m, with the caveats that currently (i.e. using the pipeline and calibration files available in August 2023 with \texttt{jwst\textunderscore1112.pmap}) the systematic uncertainties on the flux-calibrated stellar extraction for MIRI-LRS are likely on the order of 10\% (private communication, JWST Help Desk).  To verify our data reduction was not inducing the excess, we repeated our analysis on three calibrator stars, none of which showed excess.

\subsection{WASP-39b Transmission Spectrum}
To assess the excess flux in the transmission spectrum, we used the \texttt{Eureka!} transmission spectrum from \citewm. The full reduction details are fully described in \citewm~and closely followed those of \citet{bell2023_ers}, and the key steps are briefly summarized here. A column-by-column background subtraction was run on each integration, the spectrum was extracted using variance-weighted optimal spectral extraction methods \citep{HorneOptimalExtractionAlgorithm1986}, and lightcurves were binned to a constant 0.25\,$\mu$m resolution. A \texttt{starry} \citep{LugerStarryAnalyticOccultation2019} transit model was fitted to each channel assuming the orbital parameters of \citet{datasynthesis2023} and placing a Gaussian prior on the stellar limb-darkening coefficients using \texttt{ExoTETHyS} \citep{Morello2020_aj, Morello2020_joss} models based on the Stagger grid \citep{Chiavassa2018}. The systematic noise model consisted of a linear trend in time, a linear decorrelation against changes in the spatial position and PSF-width, an exponential ramp in time, and a white noise multiplier. Lightcurves were fit using the No U-Turn Sampler from \texttt{PyMC3} \citep{pymc3}.


\subsubsection{Atmospheric Modelling of WASP-39b}\label{sec:retrievals}

Here we aim to estimate the atmospheric spectrum of WASP-39b at longer wavelengths (e.g., $\gtrsim10\mu$m) as predicted by models that have been informed by the MIRI LRS data at shorter wavelengths (e.g.,$\lesssim10\mu$m). To do this, we perform an atmospheric retrieval using Aurora \citep{Welbanks2021} on just those wavelengths of the MIRI data. Aurora computes the spectrum for a parallel-plane atmosphere in transmission geometry assuming hydrostatic equilibrium. The chemical abundances are assumed to be constant with height and parameterized with a free parameter for their individual volume mixing ratios. We use the same atmospheric model setup as in Powell 2023, with the same priors and sources for absorption cross-sections. Briefly summarized here, we use an isothermal model with inhomogeneous gray clouds and power-law hazes following the single-sector prescription in \citet{Welbanks2021}, and we fit for the volume mixing ratios of H$_2$O \citep{Rothman2010} and SO$_2$ \citep{Underwood2016}. Additionally, we fit for the reference pressure for an assumed planetary radius of R$_{\mathrm{p}}=1.279~\mathrm{R}_{\mathrm{J}}$.

The retrieved atmospheric properties are weakly constrained using this model and data, in agreement with the results from Powell et al. 2023. The retrieved abundances are $\log_{10}(\text{SO}_2)=-5.8 ^{+ 1.5 }_{- 1.3 }$ and $\log_{10}(\text{H}_2\text{O})=-4.1 ^{+ 2.5 }_{- 4.6 }$. The isothermal temperature retrieved is $\text{T}=705.7 ^{+ 224.3 }_{- 137.4 }$, while the cloud and haze properties are weakly constrained to a power-law slope of $\gamma= -4.2 ^{+ 1.7 }_{- 6.8 }$ with enhancement factor of $\log_{10}(a)=7.2 ^{+ 1.9 }_{- 3.8 }$, and a gray cloud deck at a pressure of $\log_{10}\text{P}_{\text{cloud}}=-2.0 ^{+ 2.5 }_{- 3.2 }$, both covering a fraction of $\phi_\text{clouds+hazes}=0.7 ^{+ 0.2 }_{- 0.3 }$ of the terminator. 

After performing the retrieval, we randomly sample 200 equally weighted samples from the retrieved posterior distribution from fitting the model to the observations below 10$\mu$m and compute their associated spectrum from 5 to 12$\mu$m at a constant resolution of 10,000. These 200 spectra are then used to compute a median spectrum and 1$\sigma$ and 2$\sigma$ confidence intervals, shown alongside the observations in Figure \ref{fig:stellar}, middle. Then, the inferred median spectrum was binned to the resolution of the data assuming a top-hat response function. This binned spectrum is used as the reference model for the remainder of this work.  

\subsection{Calculating Excess Flux from the Transmission Spectrum}\label{sec:excess_flux_calc}
We created an empirical excess flux model assuming that the lack of agreement between the transmission spectrum and the model is due entirely to excess flux i.e., the retrieved spectrum (Figure  \ref{fig:stellar}, middle) is the true, geometric transit spectrum for the planet.  If there is a background source contaminating the transit, the transit will be diluted  by a dilution factor as a function of wavelength, $D(\lambda)$, where $D$ can be calculated as:

\begin{equation}
   D(\lambda)= \frac{d_{obs}(\lambda)}{d_{geo}(\lambda)}=\frac{F_*(\lambda)}{F_*(\lambda)+F_d(\lambda)}
\end{equation}
where $\lambda$ is wavelength,  $F_*(\lambda)$ is the stellar flux, $F_d(\lambda)$ is the excess flux or the disk flux, $d_{obs}(\lambda)$ is the observed transit depth and $d_{geo}(\lambda)$ is the real, geometric transit depth.  Using the retrieved model as $d_{geo}(\lambda)$, the calculated transit spectrum as $d_{obs}(\lambda)$, we can easily calculate $D$; using the photospheric model from  Section \ref{sec:stellarsed} as $F_*$, we can also simply solve for $F_d(\lambda)$.  In Figure \ref{fig:stellar} (bottom), we plot the empirical excess flux calculated from the dilution factor, which is $1/D$, in green.  We also plot the excess flux calculated from comparing the observed stellar flux to the photospheric model in blue. This gives us a spectrum of the contamination source as a function of wavelength.  In both cases,  at wavelengths beyond $\sim$10 $\mu$m, we calculate excess on the order of a few percent  


\subsection{Modeling the Excess Flux as a Debris Disk}



To find the basic characteristics of a debris disk that could explain the excess flux, we generate simple models of a narrow circular ring of optically thin dust grains in radiative equilibrium with the star. The flux density emitted by the disk is entirely determined by the semi-major axis of the ring (and therefore its temperature) and the amount of dust it contains. Using these two free parameters, we try models of two types: 1) dust grains that behave like blackbodies and 2) small grains composed of crystalline olivine that have a pronounced emission peak at $\sim$11$\mu$m. Although we could generate much more complex models that distribute the dust in a broad ring and have multiple compositional components,  even a very simple model can explain the data. For both model types, we assume a stellar luminosity of 0.63 L$_\odot$ and T$_{\rm
eff}$=5400~K, and we test fits to both methods of calculating the excess flux (from the stellar
and transmission spectra as described above) while also constraining the model to be consistent
with the W4 upper limit. WISE W3-band shows a small excess over the photosphere and over the JWST spectra (1.9$\sigma$) that we do not attempt to model. We use the least squares method to find best fit models by minimizing
the chi-squared goodness-of-fit metric. Because the W4 point is a 95\% confidence upper limit, we fit models constrained by 30 different W4 values spanning the range of possibilities from the 99\% confidence upper limit (1.94 mJy) down to a value equal to an excess equal to the stellar photosphere (0.69 mJy).  Overall, we used 30 different adopted W4 points, evenly spaced in flux.  We calculate formal uncertainties on the two parameters, using formulae for the variation in chi-squared as described e.g. in
\citet{BevingtonDataReductionError2003}, but because of the large uncertainties and short wavelength coverage and existence of only a W4 upper limit, the 3$\sigma$ ranges are quite large.


\section{Results}

\begin{figure}
\centering
\includegraphics[width=0.98\textwidth]{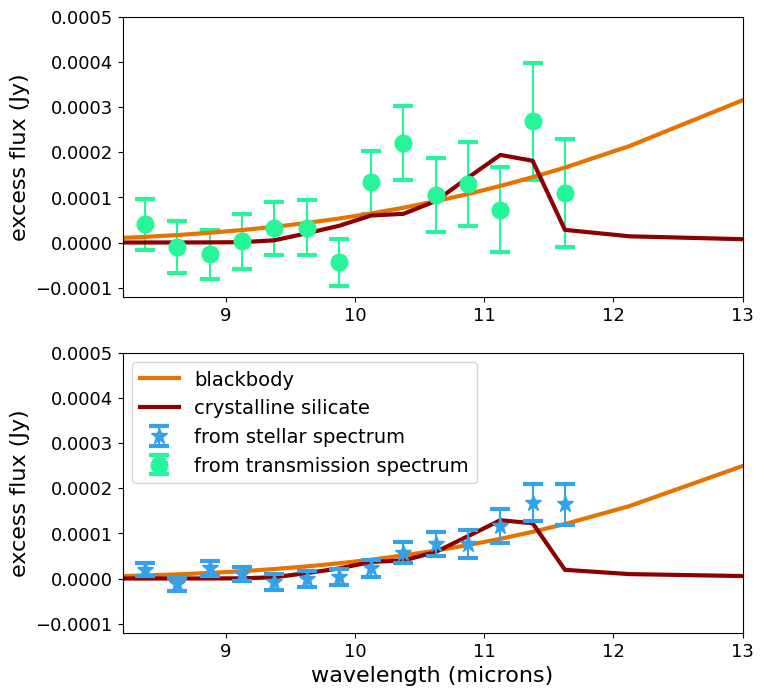}
\caption{The best fits to the excess flux as calculated from the transmission spectrum (top) or from the stellar spectrum (bottom).   We fit a crystalline silicate feature and a blackbody curve to both spectra separately.  These plots assume the 3$\sigma$ upperlimit to the WISE/W4 flux. }
\label{fig:sed_fits}
\end{figure}

The excess derived from the transmission spectrum has larger uncertainties and is compatible
with a larger range of disk models including those that best fit the data from the stellar
spectrum method.  For blackbody grains, the best fits (shown in orange solid lines in Figure
\ref{fig:sed_fits}, parameters in Table \ref{tab:modeldisksparameters}) have distances from the
star of $\sim$3.4 and 4 au (dust temperatures $\sim$135 and 124~K), for the transmission and
stellar spectrum respectively. The uncertainties on the flux densities imply that a wide range
of dust temperatures are actually allowed by the data; a 3$\sigma$ lower limit of about 1.3 au
is set by the lack of excess shorter than 10$\mu$m. The effect of including the W4 upper limit is to set an
outer radius for the disk, because cooler grains further from the star produce more 25 $\mu$m
emission. 
The maximum distance allowed by the W4 upper limit is about 8 au.

For crystalline silicate grains, the best fit temperatures are somewhat cooler, with best fit
disk radii of 13 or 16 au for the transit and stellar spectra, respectively, and a minimum ring
size of 4 au and a maximum ring size of about 30 au.

One important difference between the models is what they predict for the 12--25 $\mu$m
emission, because the blackbody models continue to rise monotonically past 12 $\mu$m. Cooler
dust is allowed in the silicate models because of the emission peaks are fairly narrow.

All best fit disk models are relatively bright, with L$_{IR}$/L$_*>3\times$
10$^{-4}$. L$_{IR}$/L$_*$ is the modeled disk luminosity in the IR integrated over wavelengths out to
1~mm divided by the total stellar luminosity.  For debris disks detected in other systems,
L$_{IR}$/L$_*$ ranges from $\sim$10$^{-5}$ to $\sim$10$^{-2}$  over a range of temperatures usually $<$100 K. Below that systems are too faint
too detect; above that are protoplanetary disks.  In comparison, the Asteroid Belt has
L$_{IR}$/L$_*\sim$10$^{-7}$ \citep{HughesDebrisDisksStructure2018}.  While we do not measure
the mass directly, as measurements out to 12 $\mu$m are only sensitive to small dust grains, this implies the disk is
fairly massive. Micron-sized dust in mature debris disks come from the collisions of
planetesimals, so a detection of dust requires the presence of larger planetesimals in the
debris disk \citep[and sources therein]{WyattEvolutionDebrisDisks2008}.  The mass in dust can
be esimated from L$_{IR}$/L$_*$, given the typical dust size and density and the simple
assumption that the grains absorb in proportion to their geometric cross sections
\citep{ChenPossibleMassiveAsteroid2001}. For grains to act like blackbodies, they must be larger than the
observed wavelength, so we can assume a size of 50 $\mu$m and a density typical of slightly
porous silicates of 2.7 g~cm$^{-3}$. For a ring radius of 4 au, the mass is $\sim$8$\times10^{20}$
kg, or similar to the most massive asteroid in the solar system.  The original mass in parent
bodies would presumably be larger than this, as there would actually be dust of a range of
sizes from 50 $\mu$m on up, which if described as a power law puts most of the mass in the
largest bodies.  The crystalline silicate model requires small grains, much smaller than the
wavelength of observations, say 0.5 $\mu$m. The mass in small grains is an order of magnitude
lower, $\sim$4$\times10^{19}$ kg for these models.

\begin{figure*}
\centering
\includegraphics[width=0.98\textwidth]{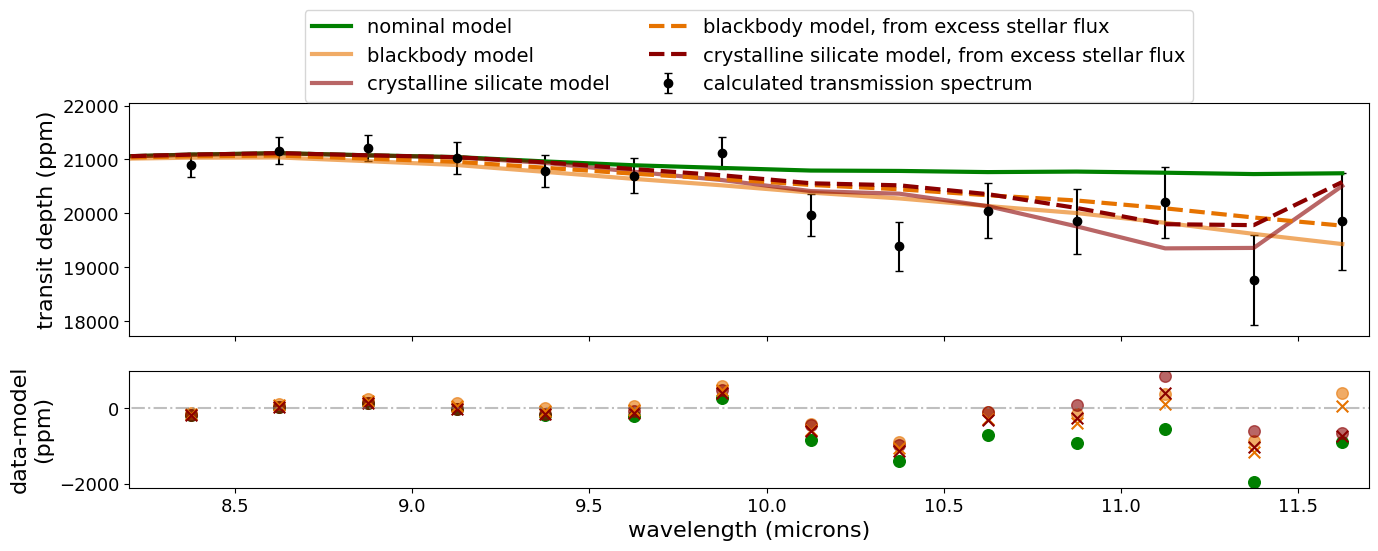}
\caption{(top) WASP-39b MIRI observed transmission spectrum along with its model retrieval and the model retrievals adjusted by the dilution factor induced by our best debris disk models.  (bottom) The residuals between the data and the models.  The inclusion of the debris disk decreases the residuals.  }
\label{fig:data_v_model}
\end{figure*}

Despite our simple models, we found that we could successfully model the data as excess flux arising in a disk. All of our models result in a better model fit to the data at wavelengths $>$9 $\mu$m (as measured by $\chi^2_\nu$) than assuming no excess (Table \ref{tab:modeldisksparameters}).  This would propagate to the exoplanet transmission spectrum, too; allowing for even these simple debris disk models results in a better fit.  These disk models result in the transmission spectra and residuals shown in Figure \ref{fig:data_v_model}.  Given the systematic uncertainties, we cannot conclusively prove that the dip in transmission spectrum or the excess flux in the stellar spectrum is caused by a debris disk, but we can clearly show that a debris disk is a plausible --- and as of now, the only --- explanation. Photometric observations at longer wavelengths could confirm or reject this hypothesis.

\begin{deluxetable*}{lrrr}
\tablecaption{Best Fit Debris Disk Model Parameters\label{tab:modeldisksparameters}}
\tablehead{%
    \colhead{Model Label} & \colhead{disk radius (au)} & \colhead{L$_{IR}$/L$_*$} 
    & \colhead{$\chi^2_\nu$} 
    }
\startdata
Blackbody Fit to Transmission Spectrum Excess & 3.4 & 8.6$\times$10$^{-4}$   & 0.80 \\
Crystalline Silicate Fit to Transmission Spectrum Excess & 12.6 & 3.4$\times$10$^{-4}$  & 0.84 \\
\hline
No Excess Flux Fit to Transmission Spectrum Excess & \nodata & \nodata  & 2.3 \\
\hline
\hline
Blackbody Fit to Stellar Spectrum Excess & 4.0 & 9.6$\times$10$^{-4}$   & 2.0 \\
Crystalline Silicate Fit to Stellar Spectrum Excess & 15.7 & 3.4$\times$10$^{-4}$ & 2.1  \\
\hline
No Excess Flux Fit to Stellar Spectrum Excess & \nodata & \nodata  & 5.1 \\
\enddata
\tablecomments{The parameters for our four best fit models (shown in Figure \ref{fig:sed_fits}) assuming the 3$\sigma$ upperlimit to the WISE/W4 flux. The $\chi^2_\nu$ for the No Excess Flux Fits were calculated with a flat line at 0 Jy.}
\end{deluxetable*}



%

\section{Discussion}

\subsection{The Effect of a Debris Disk on the C/O ratio and Metallicity of WASP-39b}
C/O ratio and metallicity can, in theory, be used to trace the formation location of an exoplanet \citep{ObergEffectsSnowlinesPlanetary2011, EspinozaMetalEnrichmentLeads2017}.  However, a low C/O ratio cannot uniquely trace a formation location, largely because a low C/O ratio can be reached via a number of paths, including forming within the H$_2$O ice line or significant planetesimal impacts. Characterizing the planetesimal population in a debris disk would allow us to better understand planet formation, because it will help break this degeneracy between formation location and planetesimal impacts.

Analysis of near-IR JWST data for WASP-39b \citep{RustamkulovEarlyReleaseScience2023, AldersonEarlyReleaseScience2023, FeinsteinEarlyReleaseScience2023, AhrerEarlyReleaseScience2023,Welbanks2024} showed that the system likely has a sub-stellar C/O ratio and super-solar metallicity.  \citet{FeinsteinEarlyReleaseScience2023} speculated that one way these could have occurred was by significant planetesimal accretion after the planet formed with planetestimals from 2-10 au.  The potential debris disk described by several of our best fit models (i.e. large and between 4-10 au) would be consistent with the source of planetesimals needed to lower WASP-39b's C/O ratio and raise its metallicity.  

\subsection{Potential Contamination of MIRI Data for Other Exoplanets}
\begin{figure}
\includegraphics[width=0.98\textwidth]{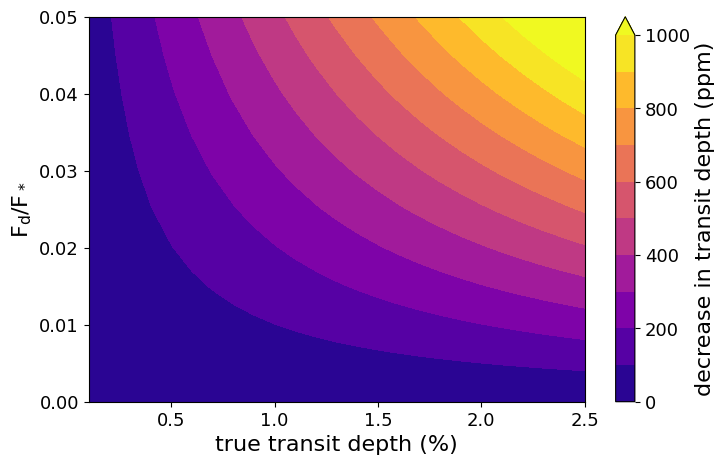}
\caption{The decrease in transit depth due to a relatively faint debris disks can be on the order of several hundred ppm. }
\label{fig:contamination}
\end{figure}

 Regardless of whether WASP-39 has a debris disk, debris disk contamination is a serious potential issue for JWST/MIRI observations of exoplanets.  In Figure \ref{fig:contamination}, we plot the magnitude of the dip induced by varying amounts of excess flux from a disk based on Equation 1. The decrease in transit depth due to a faint debris disk, even fainter than the potential disk around WASP-39, can be on the order of several hundred parts-per-million (ppm).   This is larger than the amplitude of many potentially detectable spectral features produced by exoplanets in this wavelength range.  It is also comparable to or larger than typical uncertainties in transmission spectra of exoplanets. 
 
 Since JWST/MIRI is sensitive to excesses induced by debris disks on the order of 1\%, we need to be able to measure fluxes to that level to determine whether debris disk contamination is likely to be an issue for any given system.  Unfortunately, WISE alone is not sensitive enough. For exoplanet targets observed thus far with JWST/MIRI \citep{NikolovTrExoLiSTSTransitingExoplanets2022}, the median uncertainty in WISE/W3 for the system is $\sim$2.2\%, with $\sim$85\% of targets having a W3 uncertainty less than 4.5\%, but no target had an uncertainty under 1\%.  
 
 We can also consider population statatistics to estimate what percentage of systems will have excess flux at the 1\% level. The most sensitive population surveys for warm dust use Spitzer Space Telescope or WISE data. WISE/W3 constrains this to some extent for particularly bright and warm disks. Several studies have looked at the percentage of systems with excess from debris disks. Absolute photometric calibration is the main source of uncertainty in population studies. The largest surveys, which use WISE W3 or W4 (12 or 22 $\mu$m) photometry, such as \citet{KennedyBrightEndExoZodi2013} and \citet{PatelSensitiveIdentificationWarm2014}, the typical excesses detected are $>$15\% of photospheric flux in at least one band. We then need to extrapolate to estimate the probability of disks with excess of $\sim$1\%. \citet{KennedyBrightEndExoZodi2013} estimate that 3\% of Sun-like stars will have excess greater than 1\% at 12 $\mu$m from circumstellar disks.  The LBTI is more sensitive to low levels of dust; \citet{ErtelHOSTSSurveyExozodiacal2020} showed that 20\% of systems are ``significantly dusty'' based on their flux levels at 12 $\mu$m.
 

\section{Conclusion}

We have shown that the emitted flux from circumstellar debris disks could contaminate MIRI transmission spectra of exoplanets, and may be a potential explanation for the the dip in the transit spectra seen in \citewm. We showed that the stellar spectrum also has excess flux at wavelengths longer than 10 $\mu$m, again consistent with a debris disk.  If there is a circumstellar debris disk around WASP-39b causing these features in the spectra, it is relatively large (L$_{IR}$/L$_*>$10$^{-4}$), further out than 2 au, and could be the reason why WASP-39b has a sub-stellar C/O ratio and a super-solar metallicity.  Data at longer wavelengths could confirm the presence of the disk and better constrain its properties.  

\section{Acknowledgments}
We would like to thank the anonymous referee for their helpful comments.  This work is based on observations made with the NASA/ESA/CSA JWST. The data were obtained from the Mikulski Archive for Space Telescopes at the Space Telescope Science Institute, which is operated by the Association of Universities for Research in Astronomy, Inc., under NASA contract NAS 5-03127 for JWST.  T.J.B.~acknowledges funding support from the NASA Next Generation Space Telescope Flight Investigations programme (now JWST) through WBS 411672.07.05.05.03.02.

These observations are associated with program JWST-ERS-01366. Support for program JWST-ERS-01366 was provided by NASA through a grant from the Space Telescope Science Institute.

The JWST data presented in this article were obtained from the Mikulski Archive for Space Telescopes (MAST) at the Space Telescope Science Institute. The specific observations analyzed can be accessed via \dataset[DOI]{https://doi.org/10.17909/wm2n-0j50}. 

\software{SpectRes \citep{CarnallSpectResFastSpectral2017}, NumPy \citep{oliphant2006guide, van2011numpy},    Matplotlib \citep{Hunter:2007}, astropy \citep{CollaborationAstropyCommunityPython2013}, synphot \citep{STScIDevelopmentTeamSynphotSyntheticPhotometry2018}} 


\begin{thebibliography}{}
\expandafter\ifx\csname natexlab\endcsname\relax\def\natexlab#1{#1}\fi
\providecommand{\url}[1]{\href{#1}{#1}}
\providecommand{\dodoi}[1]{doi:~\href{http://doi.org/#1}{\nolinkurl{#1}}}
\providecommand{\doeprint}[1]{\href{http://ascl.net/#1}{\nolinkurl{http://ascl.net/#1}}}
\providecommand{\doarXiv}[1]{\href{https://arxiv.org/abs/#1}{\nolinkurl{https://arxiv.org/abs/#1}}}

\bibitem[{Ahrer {et~al.}(2023)Ahrer, Stevenson, Mansfield, Moran, Brande, Morello, Murray, Nikolov, {Petit dit de la Roche}, Schlawin, Wheatley, Zieba, Batalha, Damiano, Goyal, Lendl, Lothringer, Mukherjee, Ohno, Batalha, Battley, Bean, Beatty, Benneke, {Berta-Thompson}, Carter, Cubillos, Daylan, Espinoza, Gao, Gibson, Gill, Harrington, Hu, Kreidberg, Lewis, Line, {L{\'o}pez-Morales}, Parmentier, Powell, Sing, Tsai, Wakeford, Welbanks, Alam, Alderson, Allen, Anderson, Barstow, Bayliss, Bell, Blecic, Bryant, Burleigh, Carone, Casewell, Changeat, Chubb, Crossfield, Crouzet, Decin, D{\'e}sert, Feinstein, Flagg, Fortney, Gizis, Heng, Iro, Kempton, Kendrew, Kirk, Knutson, Komacek, Lagage, Leconte, {Lustig-Yaeger}, MacDonald, Mancini, May, Mayne, Miguel, {Mikal-Evans}, Molaverdikhani, Palle, Piaulet, Rackham, Redfield, Rogers, Roy, Rustamkulov, Shkolnik, Sotzen, Taylor, Tremblin, Tucker, Turner, {de Val-Borro}, Venot, \& Zhang}]{AhrerEarlyReleaseScience2023}
Ahrer, E.-M., Stevenson, K.~B., Mansfield, M., {et~al.} 2023, Nature, 614, 653, \dodoi{10.1038/s41586-022-05590-4}

\bibitem[{Alderson {et~al.}(2023)Alderson, Wakeford, Alam, Batalha, Lothringer, Adams~Redai, Barat, Brande, Damiano, Daylan, Espinoza, Flagg, Goyal, Grant, Hu, Inglis, Lee, {Mikal-Evans}, {Ramos-Rosado}, Roy, Wallack, Batalha, Bean, Benneke, {Berta-Thompson}, Carter, Changeat, Col{\'o}n, Crossfield, D{\'e}sert, {Foreman-Mackey}, Gibson, Kreidberg, Line, {L{\'o}pez-Morales}, Molaverdikhani, Moran, Morello, Moses, Mukherjee, Schlawin, Sing, Stevenson, Taylor, Aggarwal, Ahrer, Allen, Barstow, Bell, Blecic, Casewell, Chubb, Crouzet, Cubillos, Decin, Feinstein, Fortney, Harrington, Heng, Iro, Kempton, Kirk, Knutson, Krick, Leconte, Lendl, MacDonald, Mancini, Mansfield, May, Mayne, Miguel, Nikolov, Ohno, Palle, Parmentier, {Petit dit de la Roche}, Piaulet, Powell, Rackham, Redfield, Rogers, Rustamkulov, Tan, Tremblin, Tsai, Turner, {de Val-Borro}, Venot, Welbanks, Wheatley, \& Zhang}]{AldersonEarlyReleaseScience2023}
Alderson, L., Wakeford, H.~R., Alam, M.~K., {et~al.} 2023, Nature, 614, 664, \dodoi{10.1038/s41586-022-05591-3}

\bibitem[{Allard {et~al.}(2003)Allard, Guillot, Ludwig, Hauschildt, Schweitzer, Alexander, \& Ferguson}]{AllardModelAtmospheresSpectra2003}
Allard, F., Guillot, T., Ludwig, H.-G., {et~al.} 2003, 211, 325.
\newblock \url{https://ui.adsabs.harvard.edu/abs/2003IAUS..211..325A}

\bibitem[{Allard {et~al.}(2012)Allard, Homeier, \& Freytag}]{AllardModelsVerylowmassStars2012}
Allard, F., Homeier, D., \& Freytag, B. 2012, Philosophical Transactions of the Royal Society of London Series A, 370, 2765, \dodoi{10.1098/rsta.2011.0269}

\bibitem[{Aller {et~al.}(2020)Aller, {Lillo-Box}, Jones, Miranda, \& Barcel{\'o}~Forteza}]{Aller2020tpfplotter}
Aller, A., {Lillo-Box}, J., Jones, D., Miranda, L.~F., \& Barcel{\'o}~Forteza, S. 2020, 635, A128, \dodoi{10.1051/0004-6361/201937118}

\bibitem[{{Bailer-Jones} {et~al.}(2021){Bailer-Jones}, Rybizki, Fouesneau, Demleitner, \& Andrae}]{Bailer-JonesEstimatingDistancesParallaxes2021}
{Bailer-Jones}, C. A.~L., Rybizki, J., Fouesneau, M., Demleitner, M., \& Andrae, R. 2021, The Astronomical Journal, 161, 147, \dodoi{10.3847/1538-3881/abd806}

\bibitem[{Bell {et~al.}(2022)Bell, Ahrer, Brande, Carter, Feinstein, Caloca, Mansfield, Zieba, Piaulet, Benneke, Filippazzo, May, Roy, Kreidberg, \& Stevenson}]{BellEurekaEndtoEndPipeline2022}
Bell, T., Ahrer, E.-M., Brande, J., {et~al.} 2022, The Journal of Open Source Software, 7, 4503, \dodoi{10.21105/joss.04503}

\bibitem[{Bell {et~al.}(2023)Bell, Crouzet, \& {et al.}}]{bell2023_ers}
Bell, T.~J., Crouzet, N., \& {et al.} 2023, in prep.

\bibitem[{Bevington \& Robinson(2003)}]{BevingtonDataReductionError2003}
Bevington, P.~R., \& Robinson, D.~K. 2003.
\newblock \url{https://ui.adsabs.harvard.edu/abs/2003drea.book.....B}

\bibitem[{Bushouse {et~al.}(2022)Bushouse, Eisenhamer, Dencheva, Davies, Greenfield, Morrison, Hodge, Simon, Grumm, Droettboom, Slavich, Sosey, Pauly, Miller, Jedrzejewski, Hack, Davis, Crawford, Law, Gordon, Regan, Cara, MacDonald, Bradley, Shanahan, Jamieson, Teodoro, \& Williams}]{BushouseJWSTCalibrationPipelineVersion182_2022}
Bushouse, H., Eisenhamer, J., Dencheva, N., {et~al.} 2022, Zenodo, \dodoi{10.5281/zenodo.7314521}

\bibitem[{Bushouse {et~al.}(2023)Bushouse, Eisenhamer, Dencheva, Davies, Greenfield, Morrison, Hodge, Simon, Grumm, Droettboom, Slavich, Sosey, Pauly, Miller, Jedrzejewski, Hack, Davis, Crawford, Law, Gordon, Regan, Cara, MacDonald, Bradley, Shanahan, Jamieson, Teodoro, \& Williams}]{BushouseJWSTCalibrationPipelineVersion111_2023}
---. 2023, Zenodo, \dodoi{10.5281/zenodo.8067394}

\bibitem[{Carnall(2017)}]{CarnallSpectResFastSpectral2017}
Carnall, A.~C. 2017, arXiv:1705.05165 [astro-ph].
\newblock \doeprint{1705.05165}

\bibitem[{Carter {et~al.}(2023)Carter, May, \& {et al.}}]{datasynthesis2023}
Carter, A.~L., May, E.~M., \& {et al.} 2023, in prep.

\bibitem[{Chen \& Jura(2001)}]{ChenPossibleMassiveAsteroid2001}
Chen, C.~H., \& Jura, M. 2001, The Astrophysical Journal Letters, 560, L171, \dodoi{10.1086/324057}

\bibitem[{Chen {et~al.}(2014)Chen, Mittal, Kuchner, Forrest, Lisse, Manoj, Sargent, \& Watson}]{ChenSpitzerInfraredSpectrograph2014}
Chen, C.~H., Mittal, T., Kuchner, M., {et~al.} 2014, The Astrophysical Journal Supplement Series, 211, 25, \dodoi{10.1088/0067-0049/211/2/25}

\bibitem[{Chiavassa {et~al.}(2018)Chiavassa, Casagrande, Collet, Magic, Bigot, Th{\'e}venin, \& Asplund}]{Chiavassa2018}
Chiavassa, A., Casagrande, L., Collet, R., {et~al.} 2018, antike und abendland, 611, A11, \dodoi{10.1051/0004-6361/201732147}

\bibitem[{Collaboration {et~al.}(2013)Collaboration, Robitaille, Tollerud, Greenfield, Droettboom, Bray, Aldcroft, Davis, Ginsburg, {Price-Whelan}, Kerzendorf, Conley, Crighton, Barbary, Muna, Ferguson, Grollier, Parikh, Nair, Unther, Deil, Woillez, Conseil, Kramer, Turner, Singer, Fox, Weaver, Zabalza, Edwards, Azalee~Bostroem, Burke, Casey, Crawford, Dencheva, Ely, Jenness, Labrie, Lim, Pierfederici, Pontzen, Ptak, Refsdal, Servillat, \& Streicher}]{CollaborationAstropyCommunityPython2013}
Collaboration, A., Robitaille, T.~P., Tollerud, E.~J., {et~al.} 2013, Astronomy and Astrophysics, 558, A33, \dodoi{10.1051/0004-6361/201322068}

\bibitem[{Damiano {et~al.}(2017)Damiano, Morello, Tsiaras, Zingales, \& Tinetti}]{DamianoNearIRTransmissionSpectrum2017}
Damiano, M., Morello, G., Tsiaras, A., Zingales, T., \& Tinetti, G. 2017, The Astronomical Journal, 154, 39, \dodoi{10.3847/1538-3881/aa738b}

\bibitem[{Edwards {et~al.}(2020)Edwards, Changeat, Baeyens, Tsiaras, {Al-Refaie}, Taylor, Yip, Bieger, Blain, Gressier, Guilluy, Jaziri, Kiefer, {Modirrousta-Galian}, Morvan, Mugnai, Pluriel, Poveda, Skaf, Whiteford, Wright, Zingales, Charnay, Drossart, Leconte, Venot, Waldmann, \& Beaulieu}]{EdwardsARESWASP76Tale2020}
Edwards, B., Changeat, Q., Baeyens, R., {et~al.} 2020, The Astronomical Journal, 160, 8, \dodoi{10.3847/1538-3881/ab9225}

\bibitem[{Ertel {et~al.}(2020)Ertel, Defr{\`e}re, Hinz, Mennesson, Kennedy, Danchi, Gelino, Hill, Hoffmann, Mazoyer, Rieke, Shannon, Stapelfeldt, Spalding, Stone, Vaz, Weinberger, Willems, Absil, Arbo, Bailey, Beichman, Bryden, Downey, Durney, Esposito, Gaspar, Grenz, Haniff, Leisenring, Marion, McMahon, {Millan-Gabet}, Montoya, Morzinski, Perera, Pinna, Pott, Power, Puglisi, Roberge, Serabyn, Skemer, Su, Vaitheeswaran, \& Wyatt}]{ErtelHOSTSSurveyExozodiacal2020}
Ertel, S., Defr{\`e}re, D., Hinz, P., {et~al.} 2020, The Astronomical Journal, 159, 177, \dodoi{10.3847/1538-3881/ab7817}

\bibitem[{Espinoza {et~al.}(2017)Espinoza, Fortney, Miguel, Thorngren, \& {Murray-Clay}}]{EspinozaMetalEnrichmentLeads2017}
Espinoza, N., Fortney, J.~J., Miguel, Y., Thorngren, D., \& {Murray-Clay}, R. 2017, The Astrophysical Journal Letters, 838, L9, \dodoi{10.3847/2041-8213/aa65ca}

\bibitem[{Faedi {et~al.}(2011)Faedi, Barros, Anderson, Brown, Collier~Cameron, Pollacco, Boisse, H{\'e}brard, Lendl, Lister, Smalley, Street, Triaud, Bento, Bouchy, Butters, Enoch, Haswell, Hellier, Keenan, Miller, Moulds, Moutou, Norton, Queloz, Santerne, Simpson, Skillen, Smith, Udry, Watson, West, \& Wheatley}]{Faedi2011}
Faedi, F., Barros, S. C.~C., Anderson, D.~R., {et~al.} 2011, 531, A40, \dodoi{10.1051/0004-6361/201116671}

\bibitem[{Feinstein {et~al.}(2023)Feinstein, Radica, Welbanks, Murray, Ohno, Coulombe, Espinoza, Bean, Teske, Benneke, Line, Rustamkulov, Saba, Tsiaras, Barstow, Fortney, Gao, Knutson, MacDonald, {Mikal-Evans}, Rackham, Taylor, Parmentier, Batalha, {Berta-Thompson}, Carter, Changeat, {dos Santos}, Gibson, Goyal, Kreidberg, {L{\'o}pez-Morales}, Lothringer, Miguel, Molaverdikhani, Moran, Morello, Mukherjee, Sing, Stevenson, Wakeford, Ahrer, Alam, Alderson, Allen, Batalha, Bell, Blecic, Brande, Caceres, Casewell, Chubb, Crossfield, Crouzet, Cubillos, Decin, D{\'e}sert, Harrington, Heng, Henning, Iro, Kempton, Kendrew, Kirk, Krick, Lagage, Lendl, Mancini, Mansfield, May, Mayne, Nikolov, Palle, {Petit dit de la Roche}, Piaulet, Powell, Redfield, Rogers, Roman, Roy, Nixon, Schlawin, Tan, Tremblin, Turner, Venot, Waalkes, Wheatley, \& Zhang}]{FeinsteinEarlyReleaseScience2023}
Feinstein, A.~D., Radica, M., Welbanks, L., {et~al.} 2023, Nature, 614, 670, \dodoi{10.1038/s41586-022-05674-1}

\bibitem[{Henning(2010)}]{HenningCosmicSilicates2010}
Henning, T. 2010, Annual Review of Astronomy and Astrophysics, 48, 21, \dodoi{10.1146/annurev-astro-081309-130815}

\bibitem[{Horne(1986)}]{HorneOptimalExtractionAlgorithm1986}
Horne, K. 1986, Publications of the Astronomical Society of the Pacific, 98, 609, \dodoi{10.1086/131801}

\bibitem[{Hughes {et~al.}(2018)Hughes, Duch{\^e}ne, \& Matthews}]{HughesDebrisDisksStructure2018}
Hughes, A.~M., Duch{\^e}ne, G., \& Matthews, B.~C. 2018, Annual Review of Astronomy and Astrophysics, 56, 541, \dodoi{10.1146/annurev-astro-081817-052035}

\bibitem[{Hunter(2007)}]{Hunter:2007}
Hunter, J.~D. 2007, Computing in Science \& Engineering, 9, 90, \dodoi{10.1109/MCSE.2007.55}

\bibitem[{Kendrew {et~al.}(2015)Kendrew, Scheithauer, Bouchet, Amiaux, Azzollini, Bouwman, Chen, Dubreuil, Fischer, Glasse, Greene, Lagage, Lahuis, Ronayette, Wright, \& Wright}]{KendrewMidInfraredInstrumentJames2015}
Kendrew, S., Scheithauer, S., Bouchet, P., {et~al.} 2015, Publications of the Astronomical Society of the Pacific, 127, 623, \dodoi{10.1086/682255}

\bibitem[{Kennedy \& Wyatt(2013)}]{KennedyBrightEndExoZodi2013}
Kennedy, G.~M., \& Wyatt, M.~C. 2013, Monthly Notices of the Royal Astronomical Society, 433, 2334, \dodoi{10.1093/mnras/stt900}

\bibitem[{Kipping \& Tinetti(2010)}]{KippingNightsidePollutionExoplanet2010}
Kipping, D.~M., \& Tinetti, G. 2010, Monthly Notices of the Royal Astronomical Society, 407, 2589, \dodoi{10.1111/j.1365-2966.2010.17094.x}

\bibitem[{Kostogryz {et~al.}(2023)Kostogryz, Shapiro, Witzke, Grant, Wakeford, Stevenson, Solanki, \& Gizon}]{KostogryzMPSATLASLibraryStellar2023}
Kostogryz, N., Shapiro, A.~I., Witzke, V., {et~al.} 2023, Research Notes of the American Astronomical Society, 7, 39, \dodoi{10.3847/2515-5172/acc180}

\bibitem[{Kreidberg(2018)}]{KreidbergExoplanetAtmosphereMeasurements2018}
Kreidberg, L. 2018, in Handbook of Exoplanets, ed. H.~J. Deeg \& J.~A. Belmonte No. 100, 100, \dodoi{10.1007/978-3-319-55333-7_100}

\bibitem[{Luger {et~al.}(2019)Luger, Agol, {Foreman-Mackey}, Fleming, {Lustig-Yaeger}, \& Deitrick}]{LugerStarryAnalyticOccultation2019}
Luger, R., Agol, E., {Foreman-Mackey}, D., {et~al.} 2019, The Astronomical Journal, 157, 64, \dodoi{10.3847/1538-3881/aae8e5}

\bibitem[{Mancini {et~al.}(2018)Mancini, Esposito, Covino, Southworth, Biazzo, Bruni, Ciceri, Evans, Lanza, Poretti, Sarkis, Smith, Brogi, Affer, Benatti, Bignamini, Boccato, Bonomo, Borsa, Carleo, Claudi, Cosentino, Damasso, Desidera, Giacobbe, {Gonz{\'a}lez-{\'A}lvarez}, Gratton, Harutyunyan, Leto, Maggio, Malavolta, Maldonado, {Martinez-Fiorenzano}, Masiero, Micela, Molinari, Nascimbeni, Pagano, Pedani, Piotto, Rainer, Scandariato, Smareglia, Sozzetti, Andreuzzi, \& Henning}]{Mancini2018}
Mancini, L., Esposito, M., Covino, E., {et~al.} 2018, 613, A41, \dodoi{10.1051/0004-6361/201732234}

\bibitem[{{Martin-Lagarde} {et~al.}(2020){Martin-Lagarde}, Morello, Lagage, Gastaud, \& Cossou}]{Martin-LagardePhasecurvePollutionExoplanet2020}
{Martin-Lagarde}, M., Morello, G., Lagage, P.-O., Gastaud, R., \& Cossou, C. 2020, The Astronomical Journal, 160, 197, \dodoi{10.3847/1538-3881/abac09}

\bibitem[{Morello {et~al.}(2020{\natexlab{a}})Morello, Claret, {Martin-Lagarde}, Cossou, Tsiara, \& Lagage}]{Morello2020_joss}
Morello, G., Claret, A., {Martin-Lagarde}, M., {et~al.} 2020{\natexlab{a}}, The Journal of Open Source Software, 5, 1834, \dodoi{10.21105/joss.01834}

\bibitem[{Morello {et~al.}(2020{\natexlab{b}})Morello, Claret, {Martin-Lagarde}, Cossou, Tsiaras, \& Lagage}]{Morello2020_aj}
---. 2020{\natexlab{b}}, The Astronomical Journal, 159, 75, \dodoi{10.3847/1538-3881/ab63dc}

\bibitem[{Morello {et~al.}(2021)Morello, Zingales, {Martin-Lagarde}, Gastaud, \& Lagage}]{MorelloPhasecurvePollutionExoplanet2021}
Morello, G., Zingales, T., {Martin-Lagarde}, M., Gastaud, R., \& Lagage, P.-O. 2021, The Astronomical Journal, 161, 174, \dodoi{10.3847/1538-3881/abe048}

\bibitem[{Nikolov {et~al.}(2022)Nikolov, Kovacs, \& Martlin}]{NikolovTrExoLiSTSTransitingExoplanets2022}
Nikolov, N.~K., Kovacs, A., \& Martlin, C. 2022, Research Notes of the AAS, 6, 272, \dodoi{10.3847/2515-5172/acabc7}

\bibitem[{{\"O}berg {et~al.}(2011){\"O}berg, {Murray-Clay}, \& Bergin}]{ObergEffectsSnowlinesPlanetary2011}
{\"O}berg, K.~I., {Murray-Clay}, R., \& Bergin, E.~A. 2011, The Astrophysical Journal Letters, 743, L16, \dodoi{10.1088/2041-8205/743/1/L16}

\bibitem[{Oliphant(2006)}]{oliphant2006guide}
Oliphant, T.~E. 2006 (Trelgol Publishing USA)

\bibitem[{Patel {et~al.}(2014)Patel, Metchev, \& Heinze}]{PatelSensitiveIdentificationWarm2014}
Patel, R.~I., Metchev, S.~A., \& Heinze, A. 2014, The Astrophysical Journal Supplement Series, 212, 10, \dodoi{10.1088/0067-0049/212/1/10}

\bibitem[{Powell {et~al.}(2024)Powell, Feinstein, Lee, Zhang, Tsai, Taylor, Kirk, Bell, Barstow, Gao, Bean, Blecic, Chubb, Crossfield, Jordan, Kitzmann, Moran, Morello, Moses, Welbanks, Yang, Zhang, Ahrer, {Bello-Arufe}, Brande, Casewell, Crouzet, Cubillos, Demory, Dyrek, Flagg, Hu, Inglis, Jones, Kreidberg, {L{\'o}pez-Morales}, Lagage, Meier~Vald{\'e}s, Miguel, Parmentier, Piette, Rackham, Radica, Redfield, Stevenson, Wakeford, Aggarwal, Alam, Batalha, Batalha, Benneke, {Berta-Thompson}, Brady, Caceres, Carter, D{\'e}sert, Harrington, Iro, Line, Lothringer, MacDonald, Mancini, Molaverdikhani, Mukherjee, Nixon, Oza, Palle, Rustamkulov, Sing, Steinrueck, Venot, Wheatley, \& Yurchenko}]{PowellSulphurDioxideMidinfrared2024}
Powell, D., Feinstein, A.~D., Lee, E. K.~H., {et~al.} 2024, Nature, 1, \dodoi{10.1038/s41586-024-07040-9}

\bibitem[{Rothman {et~al.}(2010)Rothman, Gordon, Barber, Dothe, Gamache, Goldman, Perevalov, Tashkun, \& Tennyson}]{Rothman2010}
Rothman, L.~S., Gordon, I.~E., Barber, R.~J., {et~al.} 2010, 111, 2139, \dodoi{10.1016/j.jqsrt.2010.05.001}

\bibitem[{Rustamkulov {et~al.}(2023)Rustamkulov, Sing, Mukherjee, May, Kirk, Schlawin, Line, Piaulet, Carter, Batalha, Goyal, {L{\'o}pez-Morales}, Lothringer, MacDonald, Moran, Stevenson, Wakeford, Espinoza, Bean, Batalha, Benneke, {Berta-Thompson}, Crossfield, Gao, Kreidberg, Powell, Cubillos, Gibson, Leconte, Molaverdikhani, Nikolov, Parmentier, Roy, Taylor, Turner, Wheatley, Aggarwal, Ahrer, Alam, Alderson, Allen, Banerjee, Barat, Barrado, Barstow, Bell, Blecic, Brande, Casewell, Changeat, Chubb, Crouzet, Daylan, Decin, D{\'e}sert, {Mikal-Evans}, Feinstein, Flagg, Fortney, Harrington, Heng, Hong, Hu, Iro, Kataria, Kempton, Krick, Lendl, {Lillo-Box}, Louca, {Lustig-Yaeger}, Mancini, Mansfield, Mayne, Miguel, Morello, Ohno, Palle, {Petit dit de la Roche}, Rackham, Radica, {Ramos-Rosado}, Redfield, Rogers, Shkolnik, Southworth, Teske, Tremblin, Tucker, Venot, Waalkes, Welbanks, Zhang, \& Zieba}]{RustamkulovEarlyReleaseScience2023}
Rustamkulov, Z., Sing, D.~K., Mukherjee, S., {et~al.} 2023, Nature, 614, 659, \dodoi{10.1038/s41586-022-05677-y}

\bibitem[{Salvatier {et~al.}(2016)Salvatier, Wiecki, \& Fonnesbeck}]{pymc3}
Salvatier, J., Wiecki, T.~V., \& Fonnesbeck, C. 2016, PeerJ Computer Science, 2, e55, \dodoi{10.7717/peerj-cs.55}

\bibitem[{{STScI Development Team}(2018)}]{STScIDevelopmentTeamSynphotSyntheticPhotometry2018}
{STScI Development Team}. 2018, Astrophysics Source Code Library, ascl:1811.001.
\newblock \url{https://ui.adsabs.harvard.edu/abs/2018ascl.soft11001S}

\bibitem[{Underwood {et~al.}(2016)Underwood, Tennyson, Yurchenko, Huang, Schwenke, Lee, Clausen, \& Fateev}]{Underwood2016}
Underwood, D.~S., Tennyson, J., Yurchenko, S.~N., {et~al.} 2016, 459, 3890, \dodoi{10.1093/mnras/stw849}

\bibitem[{Van Der~Walt {et~al.}(2011)Van Der~Walt, Colbert, \& Varoquaux}]{van2011numpy}
Van Der~Walt, S., Colbert, S.~C., \& Varoquaux, G. 2011, Computing in Science \& Engineering, 13, 22

\bibitem[{Welbanks \& {et al.}(2024)}]{Welbanks2024}
Welbanks, L., \& {et al.} 2024, in prep

\bibitem[{Welbanks \& Madhusudhan(2021)}]{Welbanks2021}
Welbanks, L., \& Madhusudhan, N. 2021, 913, 114, \dodoi{10.3847/1538-4357/abee94}

\bibitem[{Wright {et~al.}(2010)Wright, Eisenhardt, Mainzer, Ressler, Cutri, Jarrett, Kirkpatrick, Padgett, McMillan, Skrutskie, Stanford, Cohen, Walker, Mather, Leisawitz, Gautier, McLean, Benford, Lonsdale, Blain, Mendez, Irace, Duval, Liu, Royer, Heinrichsen, Howard, Shannon, Kendall, Walsh, Larsen, Cardon, Schick, Schwalm, Abid, Fabinsky, Naes, \& Tsai}]{WrightWidefieldInfraredSurvey2010}
Wright, E.~L., Eisenhardt, P. R.~M., Mainzer, A.~K., {et~al.} 2010, The Astronomical Journal, 140, 1868, \dodoi{10.1088/0004-6256/140/6/1868}

\bibitem[{Wyatt(2008)}]{WyattEvolutionDebrisDisks2008}
Wyatt, M.~C. 2008, Annual Review of Astronomy and Astrophysics, 46, 339, \dodoi{10.1146/annurev.astro.45.051806.110525}

\end{thebibliography}

\bibliographystyle{aasjournal}

\appendix

\section{Alternative sources of contamination}
\label{app:alt_contam}

\subsection{Stellar blend}
\label{app:star_blend}

A star whose flux partly falls into the aperture used for spectral extraction is a stellar blend. It could be either a gravitationally bound companion or a chance aligned background star. 
WASP-39 is a single star \citep{Faedi2011,Mancini2018}. We used \texttt{tpfplotter} \citep{Aller2020tpfplotter} to search for other potential blends from the GAIA DR3 catalog, finding only two significantly fainter sources ($\Delta mag =$4.25 and 5.32) within $\sim$1$'$. We will show that they cannot be the cause of the observed spectroscopic feature.

\begin{figure}[!h]
\centering
\includegraphics[width=0.98\textwidth]{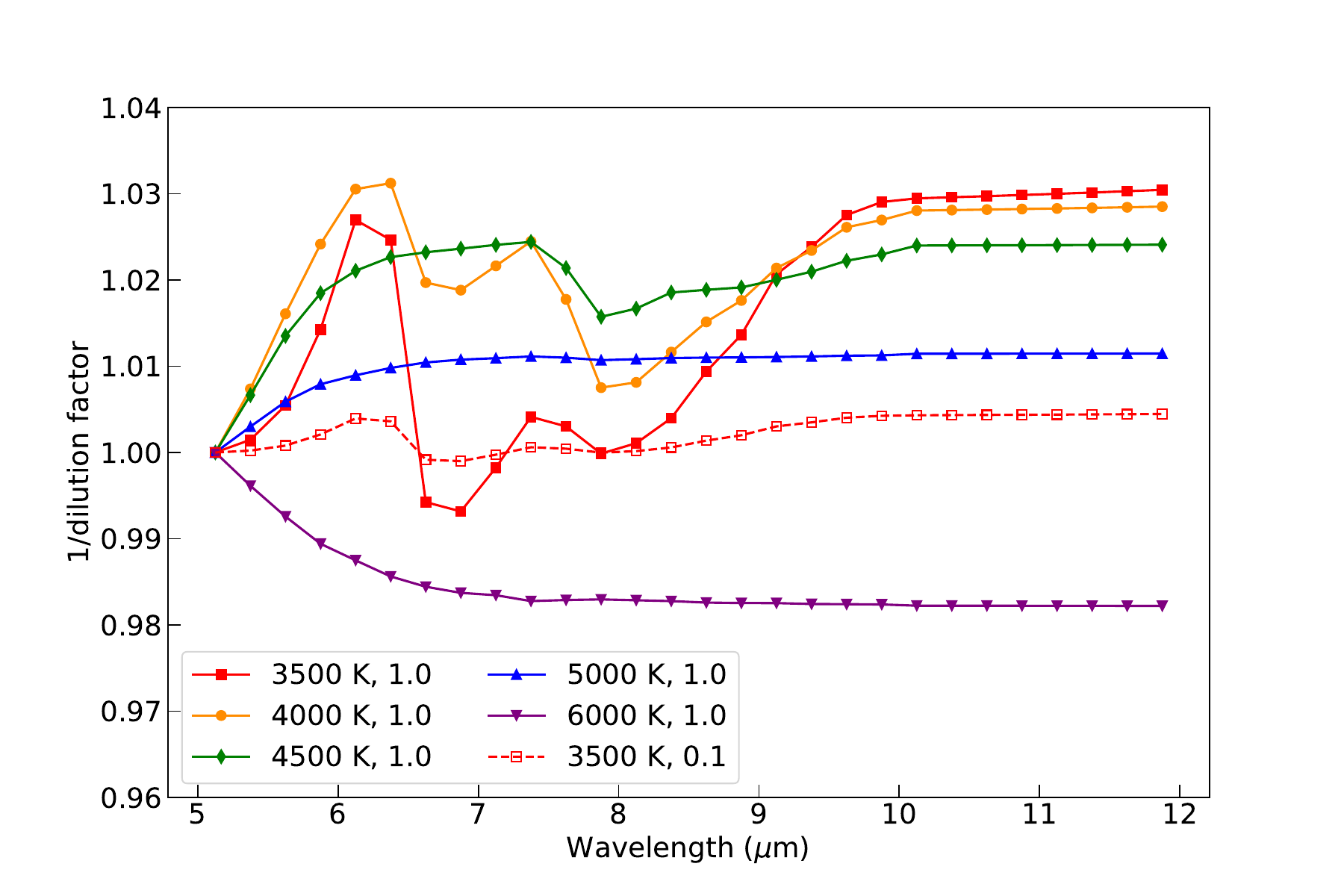}
\caption{Dilution factor on the MIRI transit spectrum for a hypothetical stellar blend with various temperatures. The labels in the legend indicate the effective temperature and flux fraction from the blended star (1.0 is the maximum possible fraction).}
\label{fig:star_dilution}
\end{figure}

Figure \ref{fig:star_dilution} reports the inverse of the dilution factor (or the excess flux), normalized to the 5--5.25\,$\mu$m bin, for a variety of stellar contaminants. We adopted the MPS-Atlas of stellar spectra \citep{KostogryzMPSATLASLibraryStellar2023}. In order to rise by a few percent at $\gtrsim$10\,$\mu$m, the contaminant should be an M dwarf or cooler star at roughly the same distance of WASP-39 b. We note that such a companion would critically affect the transmission spectrum even at shorter wavelengths.

\subsection{Planetary emission}
\label{app:planet_blend}

The emission from the planet itself may cause a similar dilution effect to that of a stellar blend, but typically smaller \citep{KippingNightsidePollutionExoplanet2010,Martin-LagardePhasecurvePollutionExoplanet2020}. However, the so-called planet self-blend effect can be significant in the infrared, where planets have their peak emission. \cite{MorelloPhasecurvePollutionExoplanet2021} estimated the possible self-contamination bias in the JWST transit spectra for a list of exoplanets, finding effects below 50 ppm for WASP-39 b. We conclude that planetary emission cannot cause the observed variation in transit depth at wavelengths $>$10\,$\mu$m.

\end{document}